\documentclass[a4paper,11pt]{article}
\usepackage{jinstpub} 
\usepackage{lineno}

\title{Diagnostic Systems in the Muon g-2 Experiment at Fermilab}

\author{M. Sorbara}
\collaboration[c]{on behalf of the Muon g-2 Collaboration}
\affiliation{Università degli Studi di Roma ``Tor Vergata'',\\
Via della Ricerca Scientifica 1, Rome, Italy}
\affiliation{Istituto Nazionale di Fisica Nucleare - Sezione di Roma Tor Vergata,\\
Via della Ricerca Scientifica 1, Rome, Italy}

\emailAdd{matteo.sorbara@roma2.infn.it}

\abstract{The muon anomalous magnetic moment, $a_\mu=\frac{g-2}{2}$, is a low-energy observable which can be both measured and computed to high precision, making it a sensitive test of the Standard Model and a probe for new physics. This anomaly was measured with a precision of $0.20$~parts per million (ppm) by the Fermilab's Muon g-2 (E989) experiment. The final goal of the E989 experiment is to reach a precision of $0.14$~ppm. The experiment is based on the measurement of the muon spin anomalous precession frequency, $\omega_a$, based on the arrival time distribution of high-energy decay positrons observed by 24 electromagnetic calorimeters, placed around the inner circumference of a $14$~m diameter storage ring, and on the precise knowledge of the storage ring magnetic field and of the beam time and space distribution. Achieving this level of precision requires strict control over systematics, which is ensured through several diagnostic devices. At the accelerator level, these devices monitor the quality of the injected beam (e.g., verifying that it has the correct momentum), while at the detector level, they track both the magnetic field and the gain of the calorimeters. In this work the devices and techniques used by the E989 experiment will be presented.}


\begin{document}
\maketitle
\flushbottom

\section{Introduction}
\label{intro}
When a charged particle travels in a region with a magnetic field, it experiences a torque, causing the spin precession around the direction of the magnetic field. The spin precession frequency is given by $\omega_s=g\frac{e}{2m}B$, where $g$ is the gyromagnetic ratio.
For a muon (and in general for an elementary lepton) the Dirac equation predicts $g=2$ at the tree level. Higher order corrections can be computed in the Standard Model framework, and defining the muon magnetic anomaly as $a_\mu = \frac{g-2}{2}$, the corrections shift the $a_\mu$ value by a factor $\sim\frac{\alpha}{2\pi}\sim 0.0012$ at first order.

The theoretical calculation of the muon magnetic anomaly includes contributions from the QED, weak interaction and two QCD-related terms: the Hadronic Vacuum Polarization (HVP) and the Hadronic Light by Light. 
The HVP term brings the highest uncertainty on the $a_\mu$ value since it cannot be computed perturbatively in the low energy region. Its calculation is based on two approaches, one uses a dispersion integral and one uses lattice QCD. 
The full calculation of $a_\mu$ has been published in 2021 from the \emph{Muon g-2 Theory Initiative} in a white paper \cite{ref:wp2020}, using the dispersive approach to compute the hadronic vacuum polarization contribution to the corrections. In 2021, the BMW collaboration presented the first result of $a_\mu$ from lattice QCD calculations with a similar uncertainty \cite{ref:bmw}. This result shows a discrepancy with the dispersive approach $a_\mu$ calculation, and a much better agreement with the experimental result. Moreover, in 2024, the CMD-3 collaboration published a result \cite{ref:cmd3} of the $e^+e^+\rightarrow\pi^+\pi^+$ cross section that, if included in the dispersive approach calculation, brings the $a_\mu$ value closer to the experimental value and to the lattice QCD estimate.

On the experimental side, $a_\mu$ was measured by the E821 Muon g-2 experiment at Brookhaven National Laboratory~\cite{ref:bnl} and, 20 years later, by the E989 experiment at Fermilab. The first result was published in 2021 using the data from the 2018 data-taking campaign (Run 1), with a precision similar to the one of the BNL experiment ~\cite{ref:run1prl}, while in 2023 results from the 2019 and 2020 (Run 2 and Run 3) campaigns were published~\cite{ref:run23}. 

\section{The Experiment}
\label{sec:experiment}
The anomalous precession frequency, ($\omega_a$), is defined as the difference between the muon's spin and cyclotron precession frequencies: 
\begin{equation}\label{eqn:master_formula}
\vec{\omega}_a=\vec{\omega}_s-\vec{\omega}_c=-\frac{e}{m}\left[a_{\mu}\vec{B} -a_{\mu}\left(\frac{\gamma}{\gamma+1}\right)\left(\vec{\beta}\cdot\vec{B}\right)\vec{\beta}-\left(a_{\mu}-\frac{1}{\gamma^2-1}\right)\frac{\vec{\beta}\times\vec{E}}{c}\right]\ ,
\end{equation}
where $\vec{E}$ is the electric field present in the region, $\vec{\beta}$ is the particle speed and $\gamma$ the Lorentz factor. For $\gamma=29.3$ and beam perpendicular to the magnetic field, both the $\vec{\beta}\times\vec{E}$ and $\vec{\beta}\cdot\vec{B}$ terms can be treated as correction terms. 

In the Muon $g-2$ experiment, a beam of positive muons, polarized above the $95\%$ level, is injected, through a magnet called \emph{inflector}, into a $14$~m diameter superconducting storage ring that produces a vertical $1.45$~T magnetic field, uniform at the ppm level. A set of pulsed non-ferric kickers inside the storage region move the muons onto the storage orbit after their injection into the ring. Four electrostatic quadrupoles (ESQ) provide the vertical focusing of the beam.

The muons' anomalous precession frequency measurement is based on the parity-violating decay of the muons in which high-energy decay positrons are emitted preferentially in the muon's spin direction. A set of 24 electromagnetic calorimeters, made of lead fluoride (PbF$_2$) \v Cerenkov crystals, are placed along the inner circumference of the storage region to measure energy and time of arrival of the decay positrons. The time distribution of the high-energy positrons can be fitted to a function, given at leading order by:
\begin{equation} \label{eqn:fitFunc}
N(t) = N_0 e^{-t/\gamma\tau}\left[1+A\cos(\omega_a t+\varphi)\right]\ ,
\end{equation}
where $N_0$ is the normalization factor, $\tau$ is the muon lifetime boosted by the Lorentz factor $\gamma$, $\varphi$ is an initial spin phase and $A$ is the decay asymmetry, that quantifies the correlation between the decay positron momentum and the muon spin directions. The value of $\omega_a$ can be extracted from this fit. Additional terms to describe beam dynamics effects are included in the full analysis (see ref. \cite{ref:run1prl} and references therein for details). 

The magnetic field intensity in equation \ref{eqn:master_formula} is averaged over the azimuth of the storage ring and weighted with the beam distribution measured using two tracking detectors.

\section{Results}
The first result of the Fermilab Muon g-2 experiment, based on 5\% of the total collected data (Run 1), was published in April 2021. It provided a confirmation of the measurement obtained by the BNL experiment two decades earlier with a similar uncertainty. 

\begin{figure}[h]
    \centering
    \includegraphics[width=0.7\textwidth]{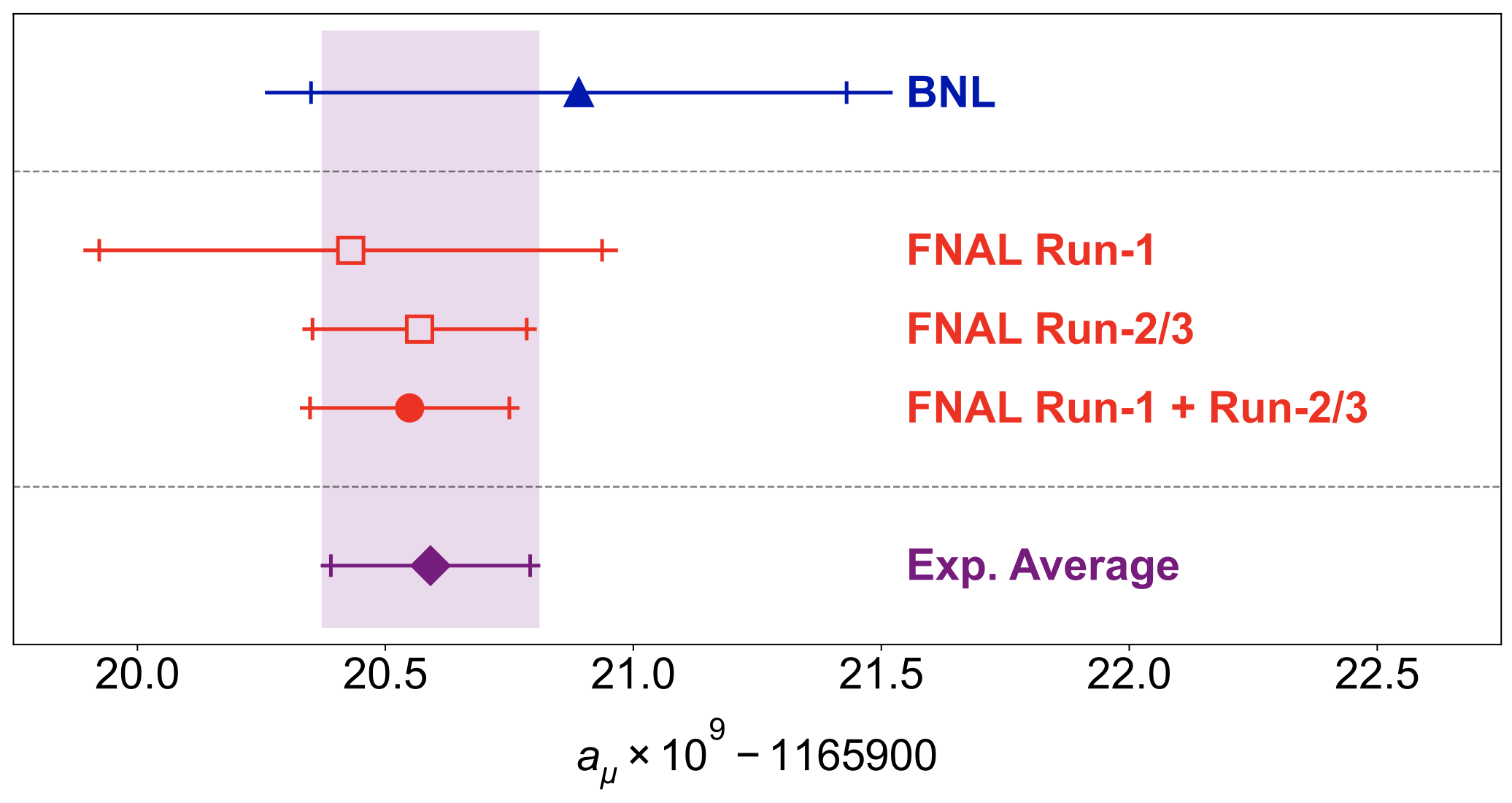}
    \caption{Summary of the Muon $g-2$ results from Run 1, Run 2, and Run 3. The experimental results are in agreement with BNL.}
    \label{fig:g2_results}
\end{figure}

On August 10, 2023, a new result incorporating data from Run 2 and Run 3 was released. This result showed excellent agreement with both the BNL and the previously published Run 1 results. The statistical uncertainty was reduced by a factor of 2.2 due to the increase in the collected data, while upgrades in the experimental hardware reduced some of the systematic effects, improving the overall precision of the experiment. 
Figure \ref{fig:g2_results} shows the results from BNL and Fermilab Run 1, 2 and 3, together with the combined average:
$$
a_\mu^{\text{ExpAvg}} = 116\,592\,059(22) \times 10^{-11} \quad [0.19 \text{ ppm}]
$$

\section{Diagnostic Systems in the Experiment}
To achieve the goal precision on $a_\mu$, several diagnostic devices are used to continuously monitor various experimental parameters during data-taking periods, providing essential information on the stored muon beam and the performance of the detectors. The monitoring ensures the stability and accuracy of the measurement, reducing systematic uncertainties and allowing real-time adjustments to optimize the experiment.

\subsection{Beam Injection Monitors}
The beam injection monitoring system in the Muon g-2 experiment has a key role in ensuring that the injected muon beam has the right conditions for optimal storage, thus increasing the number of measured muons. It is made of two components: the T0 detector and the Inflector Beam Monitoring Station (IBMS).

The T0 detector consists of a plastic scintillator slab read out by two photomultiplier tubes (PMTs). It is positioned outside the vacuum at the interface between the accelerator beamline and the entrance to the g-2 storage ring. The T0 detector provides the injection trigger, which is crucial for synchronization with other subsystems in the experiment, such as the kickers.

The IBMS is located upstream of the inflector inside the vacuum, and consists of two separate scintillating fiber detectors, both capable of measuring the muon beam intensity in the horizontal and vertical direction. These detectors are used to measure the beam profile and position before injection into the storage ring, allowing for beam tuning and alignment corrections. This ensures that the beam is optimally positioned and focused before entering the storage ring, minimizing losses and improving the overall efficiency of the experiment.

\subsection{Beam Spatial Distribution}
To reconstruct the muon beam distribution inside the storage region, two tracking stations are installed at azimuthal angles of $180^\circ$ and $270^\circ$. Each station consists of eight modules, each equipped with 32 straw tubes arranged in a stereo pattern. The tubes are filled with a mixture of Argon and Ethane. The decay positrons trajectory is reconstructed by fitting hits along these straw tubes. An extrapolation of the reconstructed trajectory allows to estimate the decay vertex coordinates inside the storage region, thus providing a direct measurement of the beam profile without interfering with the stored muon beam. The measurement of the beam profile over time also allows to extract the beam oscillation parameters that need to be included in equation \ref{eqn:fitFunc}.

\subsection{Magnetic Field Stability}
The stability of the magnetic field is crucial, as it directly enters the calculation of $a_\mu$. The magnetic field intensity is expressed in terms of the free proton precession frequency via the relation $\hbar\omega_p=2\mu_p|\vec{B}|$. This allows to use nuclear magnetic resonance (NMR) probes that measure the free proton precession frequency $\omega_p$. This technique allows for a precise measurement of the magnetic field, ensuring a reduced systematic uncertainty in the determination of $a_\mu$.

A set of 378 NMR probes is distributed around the storage ring, positioned above and below the vacuum chamber. These fixed probes are used to track any variations in the field during the normal data-taking periods. Additionally, every three days, a dedicated device equipped with 17 NMR probes is inserted into the ring to perform a high-precision mapping of the magnetic field inside the beam storage region. This periodic mapping ensures that drifts and fluctuations in the field are corrected, significantly reducing the related systematic uncertainties.

\subsection{Calorimeters Gain}
To measure the energy and arrival time of the decay positrons, the \v Cerenkov light from the calorimeter crystals is read out by Silicon Photomultipliers (SiPMs). The SiPM gain is highly sensitive to temperature variations, leading to a systematic effect in the reconstructed positron energy. Moreover, after the beam injection, the large flux of particles temporarily reduces the calorimeters gain due to the increased current demand on the power supply. To correct for these effects, a laser calibration system monitors the SiPMs response over time in order to provide corrections to the gain variations. The calibration system consists of six different lasers configured to distribute pulses in a specific pattern to each of the 1296 SiPMs via optical fibers. The SiPMs response to each pulse is analyzed to provide a correction for the gain fluctuations. The laser stability is monitored by measuring the laser intensity at the exit of the laser and at the calorimeters; this ensures that the calibration pulses are stable over time reducing the systematic effects on the energy reconstruction.

\section{Conclusions}
The Fermilab Muon $g-2$ experiment aims to measure the muon magnetic anomaly, $a_\mu$, with a precision of $140$~ppb, improving by a factor of four the previous measurement at the BNL experiment. This level of precision is only achievable with a precise knowledge of all the systematics affecting the experimental setup. Several diagnostic devices are used during the data-taking period to measure these effects in order to reduce the associated uncertainties.
With the recently published Run 1, Run 2 and Run 3 results, together with the BNL result, the experimental average for $a_\mu$ has reached a precision of $190$~ppb. The analysis of the last dataset is ongoing and the overall precision of the experiment is expected to reach its goal of $140$~ppb uncertainty on $a_\mu$.

\acknowledgments
This work was supported in part by the US DOE, Fermilab, the Istituto Nazionale di Fisica Nucleare and the European Union Horizon 2020 research and innovation programme under the Marie Sk\l{}odowska-Curie grant agreements No. 101006726, No. 734303.







\end{document}